\documentclass[twocolumn,showpacs,preprintnumbers,amsmath,amssymb]{revtex4}
\usepackage{graphicx}
\usepackage{dcolumn}
\usepackage{bm}

\begin{document}

\title{A glass anomaly in the shear modulus of solid $^4$He}

\author{Jung-Jung Su$^{1,2}$}
\author{Matthias J. Graf$^1$}
\author{Alexander V. Balatsky$^{1,2}$}
\affiliation{$^1$Theoretical Division, Los Alamos National Laboratory, Los Alamos, New Mexico 87545, USA \\
$^2$Center for Integrated Nanotechnologies, Los Alamos National Laboratory, Los Alamos, New Mexico 87545, USA}
\date{\today}

\begin{abstract}
The shear modulus of solid $^4$He exhibits an  anomalous change of order $10 \%$  \cite{Beamish07,Beamish09} at low temperatures  that is qualitatively similar to  
the much smaller frequency change in torsional oscillator experiments.   
We propose that in solid $^4$He the stiffening of the shear modulus with decreasing temperature can be described with a glass susceptibility assuming a temperature dependent relaxation time $\tau(T)$. 
The glass susceptibility captures the freezing out of glassy degrees of freedom below a characteristic crossover temperature $T_X$. There the dynamic response of the solid satisfies $\omega \tau(T_X) \sim 1$, thus leading to an increase in the shear modulus. Within this model we  predict that the maximum change of the amplitude of the shear modulus and the height of the dissipation peak are independent of the applied frequency $\omega$.   Our calculations  also show a qualitative difference in behavior of the shear modulus depending on the temperature behavior of the glass relaxation time $\tau(T)$. These predictions can be tested by comparing the complex shear modulus with experiments at different frequencies and for different levels of disorder.
\end{abstract}

\maketitle

The low-temperature anomaly of solid helium in torsional oscillators (TO) reported by Kim and Chan \cite{Chan04} has inspired an intense search for mechanical and thermodynamic anomalous properties. The observed increase of the resonance frequency below $\sim 200$ mK was taken as evidence for the mass decoupling due to supersolidity -- a quantum solid that can sustain mass superflow without dissipation. It is now accepted that the presence of defects in solid $^4$He is required to produce supersolid signatures.
Also it has been speculated that supersolidity may occur along  dislocation lines
\cite{Shevchenko87,Boninsegni07}
or grain boundaries
\cite{Burovski05,Sasaki06,Pollet07}
in solid helium.  
Direct experimental evidence for a true phase transition into a supersolid state remains inconclusive.  To date no  definitive sign of Bose-Einstein condensation (BEC) has been seen in measurements of the mass flow \cite{Beamish06,Sasaki06,Balibar08}, the melting curve \cite{Todoshchenko07}, and the lattice structure \cite{Burns08,Blackburn07}.

The basic issue with the unambiguous identification of the BEC features arises from the presence of defects that display their own dynamics and contribute to observables in the same temperature and pressure range, where supersolid anomalies are expected. We therefore are left with the dilemma: on one hand defects are required to produce supersolidity, on the other hand defects exhibit their own dynamics. Thus, any unambiguous identification of a possible supersolid state  relies on a detailed understanding of the behavior of defects.
For that reason we consider a theoretical framework that captures the dynamics of defects in solid $^4$He  in the form of a glassy component, which makes up a small fraction of the crystal \cite{Balatsky07,Nussinov07}.  This glassy component is suggested to cause the TO and thermodynamic anomalies. Further, it is consistent with reported signatures of long equilibration times, hysteresis, and a strong dependence on growth history. The glass may be created from distributions of crystal defects forming two-level-systems (TLS) \cite{Andreev09, Korshunov09}.
To determine the detailed nature of the TLS a detailed microscopic characterization of samples is needed. Possible candidates for the microscopic realization of the TLS are {\em pinned} segments of dislocation lines \cite{Balatsky07,Nussinov07,Graf08}, which naturally occur in bulk and confined solid helium. These defects can be annealed away and hence drastically change the mechanical properties of the solid. In previous work \cite{Balatsky07,Nussinov07,Graf08,Graf09,Su10} we focused on the role of the glass on the TO and thermodynamic experiments. We demonstrated that the freezing out of defect dynamics can account for the observed anomalies. Further, we found that the predicted relaxation dynamics describes the observed low-temperature features when introducing a relaxation time $\tau(T)$ that increases with decreasing temperature $T$.

Here we consider the dynamic response of elastic properties in $^4$He crystals \cite{Paalanen81,Goodkind02,Burns93,Beamish07,Beamish09}. Very recent shear modulus measurements \cite{Beamish07,Beamish09}  reveal qualitative similarities with  the TO experiments \cite{Chan04,Rittner06,Kondo07,Aoki07,Clark07,Penzev07,Hunt09,Kim09}. In the shear modulus experiment, the solid helium is grown in between two closely spaced sound transducers. When one of the transducers applies an external strain, the other transducer measures the induced stress from which the shear modulus of the sample is deduced. The experiment thus provides  a direct measurement of the elastic response to the applied force within a broad and tunable frequency range. The frequency dependence of these elastic experiments can play an important role in determining the origin of the observed anomalies.

\begin{figure}[t]
\begin{center}
\includegraphics[width=0.8\linewidth,angle=0,keepaspectratio]{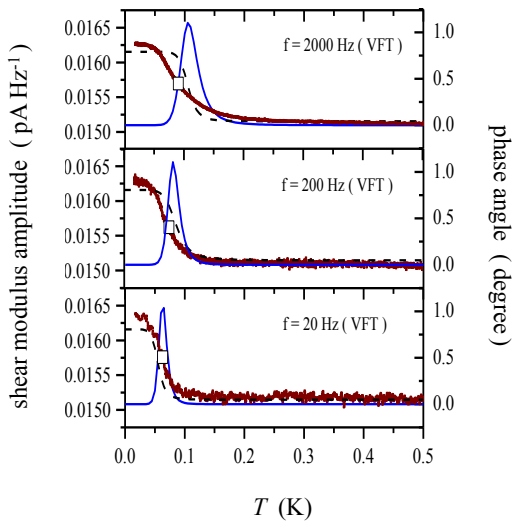}
\end{center}
\vskip-.9cm
\caption{(Color online) Experimental data and calculations (VFT) of the shear modulus vs.\ temperature. The red squares are the experimental data for the shear modulus amplitude. The black-dash line shows the calculation of the shear modulus amplitude. The blue-solid line is the prediction for the phase angle $\phi$. The calculation uses $\alpha=1.35$, $g=6.41\times 10^{-2}, \mu_0=16.2$ fA Hz$^{-1}$, 
$\tau_0=39.3$ ns, $\Delta=1.09$ K,  and $T_g=-29.8$ mK. The f=2000 Hz dataset is shifted by 0.13 fA Hz$^{-1}$ to be comparable to the 20 Hz and 200 Hz datasets.
}\label{fig:SM_VFT}
\end{figure}

In this Letter we analyze the shear modulus within the same glass framework as was introduced for the TO and specific heat experiments \cite{Balatsky07,Nussinov07,Graf08,Graf09,Su10}. Within the glass model the amplitude of the shear modulus increases (stiffens) upon lowering $T$, because defects are freezing out.  This is accompanied by a prominent dissipation peak, indicative of anelastic relaxation processes. These anomalies happen in the same temperature range where the TO anomalies were found. By studying the glass model for a complex shear modulus $\mu(\omega; T)$ we find: 
(1) The damping and amplitude of vibrations in $^4$He are controlled by the freeze-out dynamics of defects. They occur at temperatures where $\omega \tau(T) \sim 1$. In our picture the glassy contribution represents a small fraction of the total response because the glass occupies only a small fraction of the solid.   Thus we propose a theoretical description of an elastic material with a small anelastic component that is modeled by a glassy susceptibility. 
(2)  We calculate the amplitude of the shear modulus, FIG.~\ref{fig:SM_VFT}, in agreement with shear mode experiments. Within this approach we  find that the maximum of the shear modulus change and the height of the dissipation peak are independent of frequency. 
(3) We investigate  both thermal (Vogel-Fulcher-Tamman) and nonthermal (power-law) activation processes. 
(4) We predict for both processes a relation for the inverse crossover temperature $1/T_X$
 vs.\  the applied frequency $\omega$, i.e., $\omega \tau(T_X) =1$.

{\it Model -} \ 
We investigate the dynamics of the solid \cite{LL_elastic} in the presence of a back action following the logic laid out in Ref.~\cite{Nussinov07}.  Using standard response theory for elastic deformations we write the equation of motion
\begin{eqnarray} \label{EOM}
\rho \, \partial_{t}^2  u_i + \partial_j \,
\sigma_{ij}^{\rm He} \ \
= f_i^{\rm EXT} + f_i^{\rm BA} ,
\end{eqnarray}
where $\rho$ is the mass density and $u_i$ is the displacement in the $i$th direction. $f_i^{\rm EXT }$ and $f_i^{\rm BA}$ are the external force density and the back action force density in the $i$th direction.
$\sigma_{ij}^{\rm He}$ is the elastic stress tensor due to solid helium. In general, $\sigma_{ij}^{\rm He} = \lambda_{ijkl} \, u_{kl}$, with the elastic modulus tensor $ \lambda_{ijkl}$  \cite{LL_elastic}.
 The back action describes the delayed restoring force of a glass component that back-acts on the solid matrix and thus modifies the net force. 
For simplicity, we consider a homogeneous solid and set the shear wave propagation along the $z$ axis and assume that the wave polarization lies in the $x$-$y$ plane. For such a case the back action is
\begin{eqnarray} \label{fBA}
f_i^{\rm BA} = \int_{-\infty}^{t} dt' \,{\cal G}(t,t';T) \,\partial_z^2 \, u_i(t') , 
\end{eqnarray}
where $\cal G$ describes the strength of the back action on solid $^4$He and $i=x, y$. Although $f_i^{\rm BA}$ is typically much smaller than the elastic restoring force $\partial_j \, \sigma_{ij}^{\rm He}$, it is in fact this term that is responsible for the anomaly. 
The isotropic approximation is appropriate for measured polycrystalline and amorphous materials.

Applying the same approximation to the elastic
modulus tensor, $\lambda_{ijkl} = \lambda_0 \delta_{ij} \delta_{kl} + \mu_0( \delta_{ik}\delta_{jl} + \delta_{il}\delta_{jk})$,
the elastic stress tensor in Eq.~(\ref{EOM})  is finite only for orientations $j=z$ and either $k$ or $l$
equal  to $z$. With $k, l$ being interchangeable, the relevant element will
be $\lambda_{iziz}$, which gives $\mu_0$.
The fully dressed shear modulus relates the displacement to an external force, or $\rho \, \partial_{t}^2  u_i + \mu \, \partial_z^2 u_i,= f_i^{\rm EXT}$. 
Comparing this expression with Eq.~(\ref{EOM}) we obtain
\begin{eqnarray} \label{mu1}
\mu(\omega; T) =  \mu_0(T) - {\cal G} (\omega;T).
\end{eqnarray}
The back action may be described by a distribution of Debye relaxors
$
{\cal G}= \int_0^{\infty}
dt \ P(t)
\,\left[1- {1}/({1- i \omega \tau t} )\right] ,
 $
with the dimensionless parameter $t$, the normalized distribution of relaxation times $P(t)$, and the relaxation time of the glass $\tau$. 
In general $\tau(T)$ increases with decreasing $T$ and approaches infinity at the ideal glass temperature $T_g$. 
The specific form of $\tau(T)$ can change qualitatively the $T$-dependence of $\mu$ and will be discussed.
For simplicity, we choose for ${\cal G}$ the Cole-Cole distribution. Integrating over $P(t)$ yields \cite{Graf09}:
\begin{eqnarray} \label{mu}
{\cal G}(\omega;T) &=& {g_0}/{\big(1-(i \omega \, \tau) ^{\alpha}\big)} , \\
\mu(\omega; T) &=& \mu_{0} \left[ 
1 -{g}/{\big( 1-(i \omega \tau)^{\alpha}\big)} 
\right] ,
\end{eqnarray}
with the renormalized parameter $g \equiv g_0/\mu_0$, which is sample dependent. The experimental measurables are the amplitude of the shear modulus, $|\mu|$, and the phase delay between 
the input and read-out signal, $\phi \equiv  {\rm arg} \, (\mu)$;
$\phi$ measures the dissipation of the system, which is related to the inverse of the quality factor $Q^{-1} \equiv \tan \phi$.

Several interesting results follow from Eq.~(\ref{mu}). First, the change in shear modulus $\Delta \mu$ between zero and infinite relaxation time is $\Delta \mu/\mu_0 = g$, measuring the strength of the back action as well as the concentration of the TLS. Second, the peak height $\Delta \phi$ of the phase angle is proportional to $g$. When $\omega \tau =1$ (location of dissipation peak) $\Delta \phi$ becomes approximately
\begin{eqnarray}\label{PHI_G}
\Delta \phi
\approx g \, \frac{\sin(\alpha \pi/2)}{(2-g)(1-\cos(\alpha \pi/2))  } .
 \end{eqnarray}
For $g \ll 1$ and $1 < \alpha \le 2 $, Eq.~(\ref{PHI_G}) simplifies even further: $\Delta \phi \propto (2-\alpha) g \propto (2-\alpha)\Delta \mu/\mu_0$. The peak height $\Delta \phi$ depends only on the phenomenological glass parameters $\alpha$ and $g$. So both $\Delta \mu$ and $\Delta \phi$ are {\it independent of frequency}.

Finally, we comment that a glass exhibits viscoelastic properties \cite{YooDorsey09}. So far the response formulation is equivalent to a viscoelastic system with a Cole-Cole distribution of relaxation processes also known as a generalized Maxwell model 
-- parallel connections of an infinite set of Debye relaxors each with a different $\tau$. 

{\it Results -}\ We now compare our glass model calculations with the experimental shear modulus measurements by Day and Beamish \cite{Beamish07} for a transducer driven at 2000 Hz (SM2000), 200 Hz (SM200) and 20 Hz (SM20) \cite{Comment1}.

To model specific examples of  the $T$-dependent  $\tau(T)$ in  Eq.~(\ref{mu}), we consider: 
(1) the Vogel-Fulcher-Tammann (VFT) relaxation, 
and (2) the power-law (PL) relaxation, 
which represent thermal and nonthermal activation processes, respectively. 
For case (1) we assume the form:
\begin{eqnarray}
\tau(T) =
\left\{
\begin{array}{ll}
\tau_0 \,e^{\Delta/(T-T_g)} & \mbox{ for $T>T_g$} , \\
 \infty  & \mbox{ for $T \le T_g$} .
\end{array}
\right.
\end{eqnarray}
Here $\tau_0$ is the attempt time and $\Delta$ is the activation energy. The experimental data for all three frequencies are described by a single set of model parameters shown in FIG.~\ref{fig:SM_VFT}. Note that in our parameter search we did not bias  $T_g$ to be positive.  It can be either positive (real transition at $T_g$, where $\tau$ diverges) or negative $T_g$ (no real transition). Good agreement between calculations and experiments are obtained for all three frequencies for $T_g = -30$ mK. We refer to this calculation  as ``VFT$_<$''.

The calculated crossover temperature $T_X$ is slightly higher than in experiment for the SM2000 and SM200 datasets,  while slightly below the SM20 dataset. In addition, our model predicts the relative phase angle between the input and output signal, which is related to the dissipation. The phase angle exhibits a peak at $T_X$, where the amplitude of the shear modulus has a turning point. As expected $T_{X}$ decreases with decreasing $\omega$.
By setting $T_g=0$ K the VFT expression reduces to an Arrhenius rule ``VFT$_0$''. We find no distinct differences between the Arrhenius and non-Arrhenius VFT behavior (not shown).

\begin{figure}[t]
\begin{center}
\includegraphics[width=0.80\linewidth,angle=0,keepaspectratio]{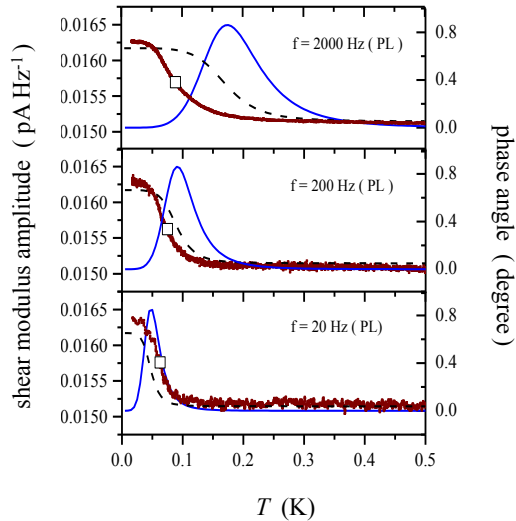}
\end{center}
\vskip-.9cm
\caption{(Color online)
Experimental data and calculations (PL) of the shear modulus vs.\ temperature. 
The red squares are the experimental data of the shear modulus amplitude. The black-dash line shows the calculation of the shear modulus amplitude. The blue-solid line shows the predicted phase delay. The model parameters are $\alpha=1.45$, $g=6.32\times 10^{-2}, \mu_0=16.2$ fA Hz$^{-1}$, 
$\tau_0=50.0$ ns, $p=3.57$, and $T_0=1.36$ K. The f=2000 Hz dataset is shifted by 0.13 fA Hz$^{-1}$ to be comparable to the 20 Hz and 200 Hz datasets.
}\label{fig:SM_PL}
\end{figure}

Next we examine case (2) of a power law for $\tau$:
\begin{eqnarray}
\tau(T) =
\tau_0 \,(T_0/T)^p ,
\end{eqnarray}
which corresponds to a freeze-out temperature at $T=0$ K. $T_0$ here is a parameter of temperature scale. The corresponding results are shown in FIG. ~\ref{fig:SM_PL}, which we refer to as ``PL$_0$''.
. The calculation is almost equally good as the one labeled VFT$_<$ for datasets SM200 and SM20. However, for  SM2000 the predicted crossover is broader and occurs at a temperature roughly 100 mK higher than in experiment.
\begin{figure}
\begin{center}
\includegraphics[width=0.95\linewidth,angle=0,keepaspectratio]{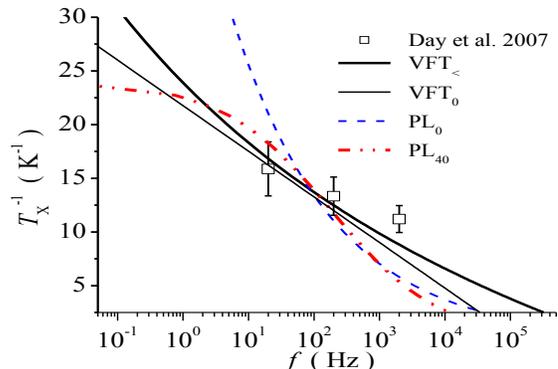}
\end{center}
\vskip-1.cm
\caption{(Color online).  Prediction for the  inverse crossover temperature vs.\ applied frequency. The black thick-solid 
(blue dashed) line is the VFT$_<$ (PL$_0$) prediction, using the same parameters as in FIG.~\ref{fig:SM_VFT} (VFT) and FIG.~\ref{fig:SM_PL} (PL). The black-thin line is the VFT prediction forcing $T_g=0$ K. The red dash-dotted line is the power-law prediction with a phase transition occurring at 40 mK (PL$_{40}$), for which we used 
$\tau=\tau_0 (|T_g|/(T-T_g))^p$ for $T>T_g$ and $\tau = \infty$ for $T \le T_g$.
}\label{fig:SM_omegavsT}
\end{figure}

To search for a possible phase transition, we study the higher and lower frequency behavior for various relaxation scenarios.
For the Cole-Cole distribution the crossover happens when $\omega \tau(T) = 1$. From that we estimate $T_{X}$ as a function of the applied frequency $f=\omega/2\pi$.

Figure~\ref{fig:SM_omegavsT} shows $1/T_X$ vs.\ $f$. The VFT$_<$ calculation gives the best description to the experimental data. The VFT$_<$ line is convex, while the Arrhenius line VFT$_0$ is straight. 
The upward curvature is typical for a VFT relaxation time with $T_g<0$, as opposed to $T_g>0$.
For comparison the power-law predictions are also shown for phase transitions occurring at 0 K (PL$_0$) and 40 mK (PL$_{40}$). For positive $T_g$ (see PL$_{40}$), we find a true freeze-out transition, indicating the arrested dynamics
for $f\to 0$ Hz. For both VFT and PL relaxation times our calculations demonstrate that in the low frequency limit the existence of a phase transition should show clear signatures of $T_X$ converging toward the glass temperature $T_g$. This behavior can serve as experimental evidence for a possible phase transition.

\begin{figure}[t]
\begin{center}
\includegraphics[width=1.0\linewidth,angle=0,keepaspectratio]{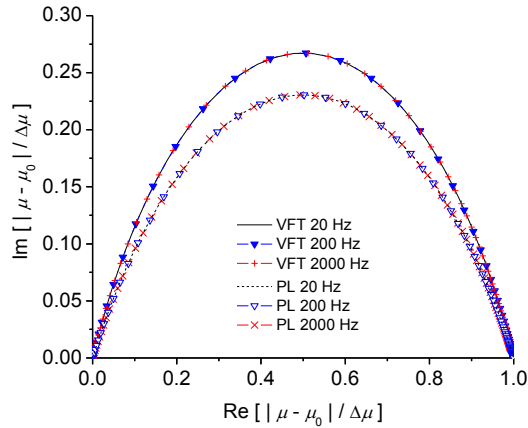}
\end{center}
\vskip-.9cm
\caption{(Color online) 
The Cole-Cole plots for relaxation times VFT$_<$ and PL$_0$ (see text). For given form of $\tau$, all different frequency curves collapse onto one single master curve reflecting that $\omega \tau$ is the only scaling parameter. The VFT$_<$ curve is higher than the PL$_0$ curve because of its smaller exponent $\alpha$. Both Cole-Cole plots show reflection symmetry about Re[$|\mu-\mu_0|/\Delta \mu$]=0.5, which is a consequence of the Cole-Cole distribution function.         
}\label{fig:CC_sppl}
\end{figure}

Finally, we show Cole-Cole plots  for both relaxation times VFT$_<$ and PL$_0$ (see FIG.~\ref{fig:CC_sppl}).  In both cases, the calculated results for all three frequencies collapse onto a single master curve demonstrating that $\omega \tau$ is the only scaling parameter in the glass model. Differences in magnitude between the VFT$_<$ and PL$_0$ calculations are due to different exponents $\alpha$ in the Cole-Cole distribution function of Debye relaxors. Of course, the use of other distribution functions may result in skewed Cole-Cole plots breaking the mirror symmetry.

In summary, we have shown that the low-temperature shear modulus anomaly of solid $^4$He can be described using the
theoretical framework of glasses. The elastic shear modulus is strongly affected by the dynamics of defects. The freezing out of defects leads to a stiffening of the solid concomitant with a peak in dissipation.  By studying the glass susceptibility due to the back action on solid helium, we find that both the amplitude change and $T$-dependence of the shear modulus are well captured by this model. 
An important consequence of the dynamic response analysis is the prediction of the dissipation or phase angle. In the proposed glass model, the peak height of the dissipation is independent of the applied frequency (in linear response) and linearly proportional to the Cole-Cole exponent $\alpha$ as well as the back action strength $g$. Since $g$ depends on the concentration of the TLS, we predict that increasing disorder will result in larger amplitude changes of the shear modulus.  
Additionally, calculations of the inverse crossover temperature vs.\ applied frequency for different thermally activated and power-law relaxation times show qualitatively different behavior. 
Finally, we find that for a positive glass temperature $T_g$  the crossover temperature $T_X$  converges toward $T_g$. This prediction can serve as a clear experimental demonstration for the existence of a true phase transition in solid $^4$He at low temperatures. We hypothesize that the freezing out of fluctuating segments of dislocation lines are the relevant excitations contributing to the reported anomalies in solid $^4$He. 
A detailed understanding of the microscopic nature of the glass and its excitations remains a pressing challenge. Future experimental characterizations may elucidate the puzzles found in the dynamics and thermodynamics of solid $^4$He.

We acknowledge fruitful discussions with Z. Nussinov, J. C. Davis, B. Hunt, E. Pratt, J. M. Goodkind, and A. Dorsey. We would 
especially like to thank J. Beamish for valuable discussions and sharing his data. This work was supported by the
 U.S.\ DOE at Los Alamos National Laboratory under contract No.~DE-AC52-06NA25396.

\end{document}